\newif\iflandscape
\newif\ifportrait
\typein[\lorp]%
{Typein "l" (for landscape, twocolumn) or "p" (for portrait, onecolumn)}
\if l\lorp \landscapetrue \else \portraittrue \fi
\newlength{\extralineskip}
\ifportrait
  \documentstyle[12pt]{article}
\fi
\iflandscape
  \documentstyle[twocolumn]{article}
\fi
\def\tr#1{{\rm tr}\kern-3pt\left[#1\right]}

\def\bea{\begin{eqnarray}}
\def\eea{\end{eqnarray}}
\def\nn{\nonumber}

\def\beq{\begin{equation}}
\def\eeq{\end{equation}}
\def\ba{\beq\new\begin{array}{c}}
\def\ea{\end{array}\eeq}
\def\be{\ba}
\def\ee{\ea}
\def\2{{1\over 2}}

\def\f{1\over}
\makeatletter
\newdimen\normalarrayskip              
\newdimen\minarrayskip                 
\normalarrayskip\baselineskip
\minarrayskip\jot
\newif\ifold             \oldtrue            \def\new{\oldfalse}
\def\arraymode{\ifold\relax\else\displaystyle\fi} 
\def\eqnumphantom{\phantom{(\theequation)}}     
\def\@arrayskip{\ifold\baselineskip\z@\lineskip\z@
     \else
     \baselineskip\minarrayskip\lineskip2\minarrayskip\fi}
\def\@arrayclassz{\ifcase \@lastchclass \@acolampacol \or
\@ampacol \or \or \or \@addamp \or
   \@acolampacol \or \@firstampfalse \@acol \fi
\edef\@preamble{\@preamble
  \ifcase \@chnum
     \hfil$\relax\arraymode\@sharp$\hfil
     \or $\relax\arraymode\@sharp$\hfil
     \or \hfil$\relax\arraymode\@sharp$\fi}}
\def\@array[#1]#2{\setbox\@arstrutbox=\hbox{\vrule
     height\arraystretch \ht\strutbox
     depth\arraystretch \dp\strutbox
     width\z@}\@mkpream{#2}\edef\@preamble{\halign \noexpand\@halignto
\bgroup \tabskip\z@ \@arstrut \@preamble \tabskip\z@ \cr}%
\let\@startpbox\@@startpbox \let\@endpbox\@@endpbox
  \if #1t\vtop \else \if#1b\vbox \else \vcenter \fi\fi
  \bgroup \let\par\relax
  \let\@sharp##\let\protect\relax
  \@arrayskip\@preamble}
\def\eqnarray{\stepcounter{equation}%
              \let\@currentlabel=\theequation
              \global\@eqnswtrue
              \global\@eqcnt\z@
              \tabskip\@centering
              \let\\=\@eqncr
              $$%
 \halign to \displaywidth\bgroup
    \eqnumphantom\@eqnsel\hskip\@centering
    $\displaystyle \tabskip\z@ {##}$%
    &\global\@eqcnt\@ne \hskip 2\arraycolsep
         $\displaystyle\arraymode{##}$\hfil
    &\global\@eqcnt\tw@ \hskip 2\arraycolsep
         $\displaystyle\tabskip\z@{##}$\hfil
         \tabskip\@centering
    &{##}\tabskip\z@\cr}
\makeatother

\def\Bf#1{\mbox{\boldmath $#1$}}

\def\balpha{{\Bf\alpha}}
\def\bbeta{{\Bf\beta}}

\def\bmu{{\Bf\mu}}

\def\blambda{{\Bf\lambda}}
\def\brho{{\Bf\rho}}

\def\f{1\over }
\def\Bf#1{\mbox{\boldmath $#1$}}

\def\balpha{{\Bf\alpha}}
\def\bbeta{{\Bf\beta}}

\def\bmu{{\Bf\mu}}

\begin{document}
\begin{titlepage}
\setcounter{footnote}0
\begin{center}
\hfill ITEP M-2/94\\
\hfill FIAN/TD-5/94\\
\hfill NBI-HE-94-27\\
\hfill hep-th/9405011\\
\vspace{0.3in}
\ifportrait
{\LARGE\bf Generalized Hirota Equations}\\{\LARGE\bf and Representation
Theory}
\\
{\LARGE\bf I. The case of $SL(2)$ and $SL_q(2)$}
\fi
\iflandscape
{\LARGE\bf Generalized Hirota Equations and Representation Theory}\\
\bigskip
{\LARGE\bf I. The case of $SL(2)$ and $SL_q(2)$}
\fi
\\[.4in]
{\Large A. Gerasimov, S. Khoroshkin\footnote{E-mail address:
khor@s43.msk.su}, D. Lebedev\footnote{E-mail address:
lebedev@vxdesy.desy.de}}\\
\bigskip {\it ITEP, Moscow, 117 259, Russia}\bigskip\\
{\Large A. Mironov\footnote{E-mail address:
mironov@fian.free.net, mironov@nbivax.nbi.dk}}\\
\bigskip {\it Theory Department,  P. N. Lebedev Physics
Institute, Leninsky prospect, 53,\\ Moscow,~117924, Russia}\\{\large
and}\\{\it Niels Bohr Institute, Blegdamsvej, 17, DK-2100 Copenhagen 0,
Denmark}
\bigskip
\\
{\Large A. Morozov\footnote{E-mail address:
morozov@vxdesy.desy.de}}\\
\bigskip {\it ITEP, Moscow, 117 259, Russia}
\end{center}
\bigskip
\bigskip

\newpage
\centerline{\bf ABSTRACT}
\begin{quotation}
This paper begins investigation of the concept of ``generalized
$\tau$-function'', defined as a generating function of
all the matrix elements of a group element $g \in G$
in a given highest-weight
representation of a universal enveloping algebra ${\cal G}$.
In the generic situation, the time-variables correspond to the
elements of maximal nilpotent subalgebras rather than Cartanian elements.
Moreover, in the case of quantum groups such $\tau$-``functions''
are not $c$-numbers but take their values in
non-commutative algebras (of functions on the quantum group $G$).
Despite all these differences
from the particular case of conventional $\tau$-functions of integrable
(KP and Toda lattice) hierarchies (which arise when
$G$ is a Kac-Moody (1-loop) algebra of level $k=1$), these
generic $\tau$-functions also satisfy bilinear Hirota-like equations,
which can be deduced from manipulations with intertwining
operators. The most
important applications of the formalism should be to
$k> 1$ Kac-Moody and multi-loop algebras, but this paper
contains only illustrative calculations for the simplest case of
ordinary (0-loop) algebra $SL(2)$ and its quantum
counterpart $SL_q(2)$, as well as for the system of fundamental
representations of $SL(n)$.
\end{quotation}
\end{titlepage}
\clearpage

\newpage

\section{Introduction}
\setcounter{footnote}{0}

The theory of $\tau$-functions is now attracting increasing attention,
because of their appearance in the role of non-perturbative partition
functions of quantum field theories. So far it was shown (see
\cite{Mamo1},\cite{Mamo2},\cite{AM}
and references therein) that $\tau$-functions
of conventional (perhaps, multicomponent) Toda-lattice hierarchy
and its reductions (like KP, KdV, MKdV etc) play this role for certain
0-dimensional theories: $c< 1$ matrix models. However, it seems
clear that this application should be much more general, though it is
also clear that this requires revision of the very concept of
$\tau$-function. The key to such generalization is provided by the
group-theoretical interpretation of conventional $\tau$-functions
\cite{J},\cite{Kac}. In this paper we shall follow the suggestion of
\cite{Leb},\cite{AM} and introduce a ``$\tau$-function'' for any
Verma module $V$ of any algebra ${\cal G}$ as a generating function for
all the matrix elements $\langle {\bf k} | g | {\bf n} \rangle_V$:
\be\label{tau}
\tau_V\{t,\bar t|g\} = \left._V
\langle {\bf 0} | \prod_{{\bf \alpha} > {\bf 0}} e_q(t_{{\bf\alpha}}
T_{{\bf\alpha}}) \ g \ \prod_{{\bf \alpha} > {\bf 0}}
e_q(\bar t_{{\bf \alpha}}T_{-{\bf \alpha}}) | {\bf 0} \rangle_V
\right.=\\
= \sum_{{{k_{\bf\alpha}\geq 0}\atop{n_{\bf\alpha}\geq 0}}}
\prod_{{\bf\alpha} > 0}
\frac{t_{\bf\alpha}^{k_{\bf\alpha}}}{[k_{\bf\alpha}]!}
\frac{\bar t_{\bf\alpha}^{n_{\bf\alpha}}}{[n_{\bf\alpha}]!}
\left._V \langle {\bf k}_{\bf\alpha} | g | {\bf n}_{\bf\alpha}
\rangle\right._V.
\ee
Here $[n] = \frac{q^n - q^{-n}}{q - q^{-1}}$,
$[n]! = [1][2]\ldots [n]$, $e_q(x) = \sum_{n\geq 0} \frac{x^n}{[n]!}$.
In the case of Lie algebras $q$-exponents are substituted by
the ordinary ones. $T_{\pm{\bf\alpha}}$ are generators of positive/negative
maximal nilpotent subalgebras $N({\cal G})$ and
$\bar N({\cal G})$ of ${\cal G}$ with
suitably chosen ordering of positive roots ${\balpha}$, and
$t_{\bf\alpha}$, $\bar t_{\bf\alpha} = t_{-{\bf\alpha}}$ are the
associated ``time-variables''. Vacuum state is annihilated by
all the positive generators:
$T_{\bf\alpha} |{\bf 0}\rangle_V = 0$ for all ${\balpha} > 0$.
Verma module $V = \left\{
|{\bf n}_{\bf\alpha} \rangle_V=
\prod_{{\bf\alpha} > 0}
T_{-{\bf\alpha}}^{n_{\bf\alpha}}| {\bf 0}\rangle_V\right\}$
is formed by the action of all the generators $T_{-{\bf\alpha}}$
for all negative roots $-{\balpha}$ from maximal nilpotent
subalgebra $\bar N({\cal G})$. Except for special circumstances
all the ${\balpha} \in N({\cal G})$ are involved,\footnote{
It can seem from (\ref{tau}) that $N({\cal G})$ and $\bar N({\cal G})$ are
already fully spanned by "the time-related" elements of ${\cal G}$ and thus the
would-be ``Grassmannian element'' $g \in G$ can be restricted
to belong to Cartan subgroup of $G$ only. This argument is, however,
misleading in the quantum group case, when matrix elements belong to
a non-commutative ``coordinate ring'' $A(G)$, while Cartanian
$g$'s span no more than its commutative sub-ring.
Our construction below does {\it not}
require $g$ to be Cartanian. See section 4 for more details.}
and since not all the $T_{-{\bf\alpha}}$'s are commuting,
the so defined $\tau$-function has nothing to do
with Hamiltonian integrability (see \cite{AM} for
detailed description of the specifics of $k=1$ Kac-Moody algebras in this
context). However, this appears to be the only property of conventional
$\tau$-functions which is not preserved by our general definition.
In particular, the goal of this paper is to explain that bilinear
(Hirota) equations are valid for our very general $\tau$-function
(\ref{tau}). In this framework they become a very general feature
of exact partition functions, essentially very close to bilinear
completeness condition.

In \cite{Kac} it was proposed to associate Hirota equations with
Casimir operators in the tensor products of irreducible representations.
However, this construction is not very straightforward in general
situation (for any representation of any algebra). Also, when
associated with non-quadratic Casimirs, it provides equations, which
are polylinear rather than bilinear in $\tau$ and are not very
convenient to deal with. Instead of the Casimir-induced construction,
we present here a very general and manifest construction, using
intertwining operators, which can be also taken as an immediate
generalization of the free-fermion approach of \cite{J} and of
projective functors of \cite{BG}, and makes
the whole story about bilinear equations very transparent and simple.
The role of the fermion is played by arbitrary vertex operator taken as an
intertwining operator into tensor product (see, for example,
\cite{BG} for classical algebras and \cite{FR} for quantum ones).
In this paper we actually describe the general scheme and
present explicit calculations only for the simplest case of $SL(2)$
and $SL_q(2)$. We emphasize, however, that despite all the
specifics of $rank\ = 1$ algebras, the construction is absolutely
general and does not rely upon any of these specific properties. Let us
repeat that the key for applicability of the same construction for
$rank\ > 1$ case is inclusion of times for {\it all} the positive
roots ${\bf\alpha}$ in the definition of $\tau$. Sometimes
(e.g. for fundamental representations of $SL(n)$) the system of bilinear
equations can be reduced to the one, involving only smaller number
of time-variables (as large as the rank of ${\cal G}$). This brings us back to
the field of Hamiltonian integrable systems.

Another aspect of the relation between our construction and conventional
$\tau$-function is that our set of bilinear equations for $SL(n)$
algebras can be splitted into that, peculiar for the Kac-Moody
$\widehat{SL(n)}|_{k=1}$ case (i.e. arising in the theory of Toda
equations) and a set of additional  constraints (playing essentially the
same role as Virasoro and $W$-constraints in matrix models),
which specifies the particular finite-parametric solution of the
form (\ref{tau}).

Section 2 describes the generic procedure to deduce bilinear
relations. Explicit examples of these equations for $G = SL_q(2)$
are considered in Section 3. Section 4 demonstrates that, despite
the possible subtlety of the notion of ``group element'' for
quantum groups, the set of solutions to bilinear equation
for $SL_q(2)$ is as big for $q\neq 1$ as it is for $q=1$.
Section 5 describes application of the general
procedure to collection of all fundamental representations of
$G = SL(N)$. The corresponding equations are very close to
those of the standard Toda-lattice hierarchy. It also contains some
comments on quantum case.

\section{From intertwining operators to bilinear equations}

We suggest the following construction in terms of intertwining
operators as a general source of
bilinear equations for the $\tau$-function (\ref{tau}). One
can easily recognize the standard free-fermion derivation of
Hirota equations for KP/Toda $\tau$-functions as a particular
example (for level $k=1$ Kac-Moody algebras $\hat G_{k=1}$,
$V$ is a fundamental representation, $W$ is the simplest fundamental
representation corresponding to the very left root of the Dynkin
diagram). Construction below involves a lot of
arbitrariness. In order to make the consideration more transparent, we
formulate our construction explicitly for finite-dimensional Lie
algebras and their $q$-counterparts. In the case of ($q$)-Affine
algebras this actually coincides with that used in \cite{FR,XXZ}.

Bilinear equations which we are going to derive
are relating $\tau$-functions (\ref{tau}) for four different
Verma modules $V,\ \hat V,\ V', \hat V'$. Given $V,\ V'$,
every allowed choice of $\hat V,\ \hat V'$ provides a separate set
of bilinear
identities. Of course, not all of these sets are actually independent
and can be parametrized by source modules $V$ and $V'$ and by a weight
of finite-dimensional representation.
Also different choices of positive root systems and their ordering
in (\ref{tau}) provides equations in somewhat different forms.
A more invariant description of the minimal set of bilinear
equations for given $G$ would be clearly interesting to find.

1. Our starting point is embedding of Verma module $\hat V$ into
the tensor product $V\otimes W$, where $W$ is some irreducible
finite-dimensional
representation of ${\cal G}$.\footnote{In the
case of Affine algebra, one should use evaluation representation -- zero charge
representation induced from finite-dimensional one -- see the
definition of vertex operator in \cite{FR}.}
Once $V$ and $W$ are specified, there is only
finite number of choices for $\hat V$.

Now we define right vertex operator of the $W$-type
as homomorphism of ${\cal G}$-modules:
\be\label{inttw}
E_R: \ \ \hat V \longrightarrow V\otimes W.
\ee

This intertwining operator
can be explicitly continued to the whole representation once this is
constructed for its vacuum
(highest-weight) state:
\be\label{vacuum}
\hat V = \left\{ | {\bf n_{\bf\alpha}} \rangle_{\hat V} =
 \prod_{{\bf\alpha}>0}
 \left(\Delta (T_{-{\bf\alpha}}\right)^{n_{\bf\alpha}}
 | {\bf 0} \rangle_{\hat V} \right\},
\ee
where comultiplication $\Delta$ provides the action of ${\cal G}$ on
the tensor product of representations\footnote{
For Lie algebra it is just $\Delta(T) = T\otimes I + I\otimes T$.},
and
\be\label{vacuum2}
|{\bf 0} \rangle_{\hat V} =
\left( \sum_{\{p_{\bf\alpha},i_{\bf\alpha}\}}
 A\{p_{\bf\alpha},i_{\bf\alpha}\}
 \left(\prod_{{\bf\alpha}>0}
 \left. (T_{-{\bf\alpha}}\right)^{p_{\bf\alpha}}\otimes
 \left. (T_{-{\bf\alpha}}\right)^{i_{\bf\alpha}}\right) \right)
| {\bf 0} \rangle_V \otimes | {\bf 0} \rangle_W.
\ee
For finite-dimensional $W$'s, this provides every
$| {\bf n_{\bf\alpha}} \rangle_{\hat V}$ in a form of {\it finite}
sums of states $| {\bf m_{\bf\alpha}} \rangle_{V}$ with
coefficients, taking values in elements of $W$.

2. The next step is to take another triple, defining left vertex
operator\footnote{
Note the change of ordering at the r.h.s., this is slightly
different from $V'\otimes W'$ in the case of quantum groups.},
\be\label{inttw2}
\bar E'_L: \ \ \hat V' \longrightarrow W' \otimes V',
\ee
such that the product $W\otimes W'$ contains {\it unit}
representation of ${\cal G}$.
The projection to this unit representation
\be\label{proj}
\pi: \ \ W\otimes W' \longrightarrow I
\ee
is explicitly provided by multiplication of any element
of $W\otimes W'$ by
\be\label{explproj}
\pi = \left._W \langle {\bf 0} | \otimes \left._{W'} \langle {\bf 0} |
\left( \sum_{\{i_{\bf\alpha},i'_{\bf\alpha}\}}
\pi\{i_{\bf\alpha},i'_{\bf\alpha}\}
 \left(\prod_{{\bf\alpha}>0}
 \left. (T_{+{\bf\alpha}}\right)^{i_{\bf\alpha}}\otimes
 \left. (T_{+{\bf\alpha}}\right)^{i'_{\bf\alpha}}\right) \right)
\right.\right.
\ee
Using this projection, if it is not occasionally orthogonal to the
image of $E\otimes E'$, one can build a new intertwining
operator
\be\label{Gamma}
\Gamma: \ \
\hat V \otimes \hat V' \stackrel{E\otimes E'}{\longrightarrow}
V \otimes W \otimes W' \otimes V'
\stackrel{I\otimes \pi \otimes I}{\longrightarrow} V \otimes V',
\ee
which possesses the property
\be\label{CRGamma}
\Gamma (g\otimes g) = (g\otimes g) \Gamma
\label{Ggg=ggG}
\ee
for any group element $g$ such that
\be
\Delta(g)=g\otimes g.
\ee

3. It now remains to take a matrix element of (\ref{Ggg=ggG})
between four states,
\be\label{CRGamma2}
\left._{V'} \langle k' | \left._V \langle k |
(g\otimes g) \Gamma | n \rangle_{\hat V}
 |n' \rangle_{\hat V'} \right.\right. =
\left._{V'} \langle k' | \left._V \langle k |
\Gamma (g\otimes g) | n \rangle_{\hat V}
|n' \rangle_{\hat V'} \right.\right.
\ee
and rewrite this identity in terms of generating functions
(\ref{tau}).

For illustrative purposes we give now an explicit example of this
calculation for the case of $SL_q(2)$. Formulas for $SL(2)$ arise
in the limit $q=1$.

\section{Explicit equations for $SL_q(2)$}

\subsection{Bilinear identities}

To begin with, fix the notations. We consider generators $T_+$,
$T_-$ and $T_0$ of $U_q(SL(2))$ with commutation relations

\be
q^{T_0} T_\pm q^{-T_0} = q^{\pm 1}T_\pm, \nn \\
\phantom. [T_+,T_-] = \frac{q^{2T_0}-q^{-2T_0}}{q - q^{-1}},
\ee
and comultiplication

\be\label{coprod}
\Delta(T_\pm) = q^{T_0} \otimes T_\pm + T_\pm \otimes q^{-T_0}, \nn \\
\Delta(q^{T_0}) = q^{T_0}\otimes q^{T_0}.
\ee
Verma module $V_\lambda$ with highest weight $\lambda$ (not obligatory
half-integer), consists of the elements
\be\label{notation1}
| n \rangle_\lambda \equiv T_-^n|0 \rangle_\lambda, \ \ n\geq0,
\ee
such that
\be\label{notation}
T_- | n \rangle_\lambda = | n+1 \rangle_\lambda, \nn \\
T_0 | n \rangle_\lambda =
 (\lambda - n) | n \rangle_\lambda , \nn \\
T_+ | n \rangle_\lambda \equiv
b_n(\lambda) | n-1 \rangle_\lambda, \nn \\
b_n(\lambda) = [n][2\lambda + 1 - n], \ \ \
 [x] \equiv \frac{q^x - q^{-x}}{q - q^{-1}}, \nn \\
|| n ||^2_\lambda \equiv \left._\lambda\langle n | n \rangle\right._\lambda
= \frac{[n]!\ \Gamma_q(2\lambda +1)}{\Gamma_q(2\lambda +1-n)}
\stackrel{\lambda \in {\bf Z}/2}{=}
\frac{[2\lambda]![n]!}{[2\lambda -n]!}.
\ee
Now,
\be\label{coprod2}
\left(\Delta(T_-)\right)^n =
q^{nT_0}\otimes T_-^n + [n]T_-q^{(n-1)T_0}\otimes T_-^{n-1}q^{-T_0}
+ \ldots + \nn \\
+ [n] T_-^{n-1}q^{T_0}\otimes T_-q^{-(n-1)T_0} + T_-^n\otimes q^{-nT_0}.
\ee

Let us manifestly derive equations (\ref{CRGamma2})
taking for $W$ an irreducible
spin-$\2$ representation of $U_q(SL(2))$. Then $\hat
V=V_{\lambda\pm {\f 2}}$, $V=V_\lambda$ and the highest weights of
$\hat V$ in $W\otimes V$ or $V\otimes W$ are\footnote{Hereafter we omit
the symbol of tensor product from the notations of the states $|+\rangle
\otimes|0\rangle_\lambda$ etc.}:

\be\label{hiweight1}
|0\rangle_{\lambda +\frac{1}{2}} =
| + \rangle | 0 \rangle_\lambda, \ \ |+\rangle\equiv|0\rangle_{\2},\\
{\rm or}\ \ \ | 0 \rangle_\lambda | + \rangle ;
\ee

\be\label{hiweight2}
|0\rangle_{\lambda - \frac{1}{2}} =
| + \rangle | 1 \rangle_\lambda -
q^{(\lambda + \frac{1}{2})}[2\lambda]| -
\rangle | 0 \rangle_\lambda, \ \ |-\rangle\equiv|1\rangle_{\2}, \\
{\rm or} \ \left(q \rightarrow q^{-1}\right) \ \ \
 | 1 \rangle_\lambda | + \rangle -
q^{-(\lambda + \frac{1}{2})}[2\lambda] | 0 \rangle_\lambda
| - \rangle.
\ee
Entire Verma module is generated by the action of $\Delta(T_-)$:

\be\label{state1}
| n \rangle_{\lambda + \frac{1}{2}} =
\left(\Delta(T_-)\right)^n |0\rangle_{\lambda + \frac{1}{2}}
\longrightarrow \nn \\
q^{n/2} \left( | + \rangle| n \rangle_\lambda +
   q^{-(\lambda +\frac{1}{2})}[n]
| - \rangle | n-1 \rangle_\lambda \right), \nn \\
{\rm or}\ \ \
q^{-n/2} \left( | n \rangle_\lambda | + \rangle +
   q^{(\lambda +\frac{1}{2})}[n]
| n-1 \rangle_\lambda | - \rangle \right);
\ee

\be\label{state2}
| n \rangle_{\lambda - \frac{1}{2}} =
\left(\Delta(T_-)\right)^n |0\rangle_{\lambda - \frac{1}{2}}
\longrightarrow \nn \\
q^{n/2} \left( | + \rangle| n+1 \rangle_\lambda +
   q^{(\lambda +\frac{1}{2})}[n-2\lambda]
| - \rangle | n \rangle_\lambda \right), \nn \\
{\rm or}\ \ \
q^{-n/2} \left( | n+1 \rangle_\lambda | + \rangle +
   q^{-(\lambda +\frac{1}{2})}[n-2\lambda]
| n \rangle_\lambda | - \rangle \right);
\ee

Step 2 to be made in accordance with our general procedure
is to project the tensor product of two different $W$'s
onto singlet state
$S = |+\rangle |-\rangle - q|-\rangle|+\rangle$:\footnote{This
state is a singlet of $U_q(SL(2))$. In the case of $U_q(GL(2))$ one
should account for the $U(1)$ non-invariance of $S$. This is the origin
of the factor  ${\rm det}_q g$ at the r.h.s. of the final equations
(\ref{qeqA}), (\ref{eqA}) and (\ref{spin1}).}

\be\label{qdet}
(A| + \rangle + B| - \rangle)\otimes
(| + \rangle C + | - \rangle D) \longrightarrow
AD-qBC.
\ee

With our choice of $W$ we can now consider
two different cases:

($A$) both $\hat
V=V_{\lambda-{\f 2}}$ and $\hat
V'=V_{\lambda'-{\f 2}}$, or

($B$) $\hat
V=V_{\lambda-{\f 2}}$ and $\hat
V'=V_{\lambda'+{\f 2}}$:

\underline{Case A}:

\be\label{projA}
| n \rangle_{\lambda - \frac{1}{2}}
| n' \rangle_{\lambda ' - \frac{1}{2}} \longrightarrow \nn \\
\longrightarrow q^{\frac{n'-n-1}{2}} \left(
[n'-2\lambda']q^{\lambda '}
| n+1 \rangle_\lambda | n' \rangle_{\lambda '} -
[n-2\lambda]q^{-\lambda}
| n \rangle_\lambda | n'+1 \rangle_{\lambda '}\right).
\ee

\underline{Case B}:

\be\label{projB}
| n \rangle_{\lambda + \frac{1}{2}}
| n' \rangle_{\lambda ' - \frac{1}{2}} \longrightarrow \nn \\
\longrightarrow q^{\frac{n'-n-1}{2}} \left(
[n'-2\lambda']q^{\lambda '}
| n \rangle_\lambda | n' \rangle_{\lambda '} -
[n]q^{+\lambda +1}
| n-1 \rangle_\lambda | n'+1 \rangle_{\lambda '}\right).
\ee

Now we proceed to the step 3. Consider any
``group element'', i.e. an element $g$ from some extension of $U_q(G)$,
which possesses
the property:

\be\label{grel}
\Delta (g)=g\otimes g,
\ee
and take matrix elements of the formula (\ref{CRGamma}):
\be\label{me1}
\phantom._{\lambda '}\langle k' |
\phantom._{\lambda}\langle k |
\left(g\otimes g \Gamma =_{\phantom.}^{\phantom.}
\Gamma g\otimes g \right)
| n \rangle_{\hat\lambda} | n' \rangle_{\hat\lambda '}.
\ee
The action of operator $\Gamma$ can be represented as:
\be\label{me2}
\Gamma | n \rangle_{\hat\lambda} | n' \rangle_{\hat\lambda'}
= \sum_{l,l'} | l \rangle_\lambda | l' \rangle_{\lambda '}
\Gamma(l,l'| n,n'),
\ee
and in these terms (\ref{me1}) turns into:
\be\label{me3}
\sum_{m,m'} \Gamma (k,k'| m,m')
\frac{|| k ||^2_\lambda
 || k' ||^2_{\lambda '}}
 {|| m ||^2_{\hat\lambda}
 || m' ||^2_{\hat\lambda '}}
\langle m | g | n \rangle_{\hat\lambda}
\langle m' | g | n' \rangle_{\hat\lambda '} = \nn \\
= \sum_{l,l'}
\langle k | g | l \rangle_\lambda
 \langle k' | g | l' \rangle_{\lambda '}
\Gamma(l,l'| n,n').
\label{bilimat}
\ee

In order to rewrite this as a difference equation, we
use our definition of $\tau$-function:

\be\label{tau2}
\tau_\lambda(t,\bar t| g) \equiv
\langle \lambda | e^{tT^+} g e^{\bar tT^-} | \lambda \rangle =
\sum_{m,n} \langle m | g | n \rangle_\lambda
\frac{t^m}{[m]!}\frac{\bar t^n}{[n]!}.
\ee
Then, one can write down the generating formula for the equation
(\ref{bilimat}), using the manifest form (\ref{projA})-(\ref{projB})
of matrix elements $\Gamma(l,l'| n,n')$:

\underline{Case A}:

\be\label{qeqA}
\sqrt{M_{\bar t}^- M_{\bar t'}^+}
\left( q^{\lambda '} D_{\bar t}^{(0)}
                   \bar t'D_{\bar t'}^{(2\lambda ')} -
  q^{-\lambda } \bar t D_{\bar t}^{(2\lambda)} D_{\bar t'}^{(0)}\right)
\tau_\lambda (t,\bar t | g)\tau_{\lambda '}(t',\bar t'| g) = \nn \\
= [2\lambda][2\lambda'](\det_q g) \sqrt{M_{t}^- M_{t'}^+}
\left( q^{-(\lambda+\frac{1}{2})}t' - q^{(\lambda' + \frac{1}{2})}t\right)
\tau_{\lambda - \frac{1}{2}}(t,\bar t | g)\tau_{\lambda ' - \frac{1}{2}}
(t',\bar t'| g).
\ee
Here
$
D_t^{(\alpha)} \equiv
\frac{q^{-\alpha}M^+_t - q^\alpha M^-_t}{(q - q^{-1})t}
$
and $M^{\pm}$ are multiplicative shift operators,
$M^{\pm}_tf(t)=f(q^{\pm 1}t)$.

\underline{Case B}:

\be\label{qeqB}
\sqrt{M_{\bar t}^- M_{\bar t'}^+}
\left( q^{\lambda '} \bar t' D_{\bar t'}^{(2\lambda ')} -
  q^{(\lambda +1)} \bar t D_{\bar t'}^{(0)}\right)
\tau_\lambda (t,\bar t | g)\tau_{\lambda '}(t',\bar t'| g) = \nn \\
= \frac{[2\lambda ']}{[2\lambda +1]} \sqrt{M_{t}^- M_{t'}^+}
\left( q^{\lambda '} tD_{t}^{(2\lambda +1)} -
  q^{\lambda} t' D_{t}^{(0)}\right)
\tau_{\lambda + \frac{1}{2}}(t,\bar t | g)\tau_{\lambda ' -
\frac{1}{2}}(t',\bar t'|g).
\ee

Let us note that the derivation of these equations can be presented in
the form more close to
the original framework of \cite{J}. Namely, we can realize operator
$\Gamma$ in component form as $E_1^R\otimes E_2^L-qE_2^RE_1^L$, where
$E_{i}$'s are components of the vertex operator\footnote{These
components $E_i$'s  certainly correspond to fermions
$\psi_i$ in \cite{J}.} (given by fixing
different vectors from $W$). Then the equation (\ref{Ggg=ggG}) can be rewritten
\be\label{J}
\left._{\hat V}
\langle 0|e_q(tT_+)\ E_1^Rg\ e_q(\bar t T_-)|0\rangle_V\right.
\cdot
\left._{\hat V'}\langle 0|e_q(tT_+)\ E_2^Lg\ e_q(\bar t
T_-)|0\rangle_V'\right.-\\-
q\left._{\hat V}\langle 0|e_q(tT_+)\ E_2^Rg\ e_q(\bar t T_-)|0\rangle_V\right.
\cdot
\left._{\hat V'}\langle 0|e_q(tT_+)\ E_1^Lg\ e_q(\bar t T_-)|0\rangle_V'\right.
=\\=
\left._{\hat V}
\langle 0|e_q(tT_+)\ gE_1^R\ e_q(\bar t T_-)|0\rangle_V\right.
\cdot
\left._{\hat V'}\langle 0|e_q(tT_+)\ gE_2^L\ e_q(\bar t
T_-)|0\rangle_V'\right.-\\-
q\left._{\hat V}\langle 0|e_q(tT_+)\ gE_2^R\ e_q(\bar t T_-)|0\rangle_V\right.
\cdot
\left._{\hat V'}\langle 0|e_q(tT_+)\ gE_1^L\ e_q(\bar t
T_-)|0\rangle_V'\right..
\ee
We can easily obtain commutation
relations of $E_i$'s with generators of algebra (see, for example,
\cite{FR}) as well as their action on vacuum states. Then, it is
immediately to
commute $E_i$'s with $q$-exponentials in the expression (\ref{J})
and to present the
result of commuting by the action of difference operators, in complete
analogy with the approach of \cite{J}. Certainly, it reproduces the results
(\ref{qeqA}) and (\ref{qeqB}).

\subsection{Solutions of bilinear identities}

Now let us consider manifest solutions of bilinear identities of the
previous subsection. We start with the case of equation (\ref{qeqA})
and the simplest ``group element'' $g=I$, which definitely
satisfies (\ref{grel}). Then,

\be\label{qsolA}
\tau_\lambda = [1 + t\bar t]^{2\lambda} \equiv
\sum_{i\geq 0} \frac{\Gamma_q(2\lambda +1)}
{\Gamma_q(2\lambda +1 - i)}\frac{(t\bar t)^i}{[i]!}
\ee
does indeed satisfy (\ref{qeqA}), since
\be\label{1}
D_{t}^{(0)} [1 + t\bar t]^{2\lambda} = [2\lambda]
   [1+ t\bar t]^{2\lambda -1}\bar t, \nn \\
tD_t^{(2\lambda)} [1 + t\bar t]^{2\lambda} = -[2\lambda]
   [1 + t \bar t]^{2\lambda - 1}.
\ee

It is instructive now to write down the classical identity,
which can be derived for $G = SL(2)$ case and is trivially obtained from
(\ref{qeqA}) in the limit $q=1$:\footnote{We use systematically
notations
$\partial$ for usual derivatives and $D$ -- for $q$-difference ones.}

\be\label{eqA}
\left(2\lambda \frac{\partial}{\partial \bar t'} -
    2\lambda' \frac{\partial}{\partial \bar t} +
 (\bar t' - \bar t)\frac{\partial^2}{\partial \bar t\partial \bar t'}
\right)
 \tau_\lambda (t,\bar t | g)\tau_{\lambda '}(t',\bar t'| g) = \nn \\
= 4\lambda\lambda ' (\det g) (t' - t)
\tau_{\lambda - \frac{1}{2}}(t,\bar t | g)\tau_{\lambda ' - \frac{1}{2}}
(t',\bar t'| g) .
\ee
For $g=I$ it has an evident solution $\tau_\lambda = (1 + t\bar t)^{2\lambda}$.

Of course the same (\ref{qsolA}) satisfies the second  equation (\ref{qeqB}),
which  in the classical limit $q=1$ looks like
\be
\phantom{fhg}\left[(\bar t'-\bar t){\partial\over\partial\bar t'}-2\lambda'
\right]
\tau_\lambda(t,\bar t|g)\tau_{\lambda '}(t',\bar t'|g)=\\=
{2\lambda '\over 2\lambda +1}\left[(t-t'){\partial\over\partial t}-
2\lambda-1\right]\tau_{\lambda+{\f 2}}(t,\bar t|g)\tau_{\lambda '-
{\f 2}}(t',\bar t'|g).
\ee
Thus we see that the same $\tau$-function satisfies differently
looking equations, arising in the cases ($A$) and ($B$). Taking other
representations to play the role of $W$, one can derive a lot of
other bilinear identities. Say, if $W$ is the
spin-$1$ representation $W$ one obtains (in the classical case)
the following identity:

\be\label{spin1}
\left[{\f 2\lambda(2\lambda-1)}{\partial^2\over\partial \bar t^2}+
{\f 2\lambda'(2\lambda'-1)}{\partial^2\over\partial \bar t'^2} -
{\over 2\lambda\lambda'}{\partial^2\over\partial \bar t\partial\bar
t'}-\right.\\
-{1\over 2\lambda\lambda'(2\lambda'-1)}(\bar t-\bar t')
{\partial\over \partial \bar t}
{\partial^2\over\partial \bar t'^2}
+{1\over 2\lambda\lambda'(2\lambda-1)}(\bar t-\bar t')
{\partial^2\over \partial \bar t^2}
{\partial\over\partial \bar t'}+\\\left.+
{1\over 4\lambda\lambda'(2\lambda '-1)(2\lambda-1)}(\bar t-\bar t')^2
{\partial^2\over \partial \bar t^2}
{\partial\over\partial \bar t'}\right]
\tau_{\lambda}(t,\bar t|g)\tau_{\lambda'}(t',\bar t'|g)=\\
=(t-t')^2(\det g)^2\tau_{\lambda-1}(t,\bar t|g)\tau_{\lambda'-1}(t',\bar t'|g).
\ee
For $SL(2)$ all new identities are, however, corollaries of the first ones.

Now let us note that the general solution of equation (\ref{eqA}) is
in fact a 3-parametrical one:

\be\label{3pfam}
\tau_{\lambda}(t,\bar t|g)=(a+bt+c\bar t+dt\bar t)^{2\lambda},
\ee
which is just explicit answer for (\ref{tau2}),
where $a$, $b$, $c$ and $d$ are elements of the $2\times 2$ matrix,
parametrizing
the group element $g$ of $GL(2)$. We shall see in the next section that
for $q\neq 1$ the set of solutions is still 3-parametric, if
considering in appropriate setting.

However, before we proceed to this issue, it deserves mentioning,
what is the relation of our equations to the Liouville one, which
is also sometimes associated with $SL(2)$ - but in a slightly
different sense.
First of all, as all the Hirota-like equations,
(\ref{eqA}) can be rewritten
as a (system of) ordinary differential equations, when expanded in
powers of $\epsilon = \frac{1}{2}(t - t')$ and
$\bar\epsilon = \frac{1}{2}(\bar t - \bar t')$. For example,
for $\lambda = \lambda'$ we obtain from (\ref{eqA}):
\be\label{expeq}
{\rm coefficient\ in\ front\ of\ }\epsilon: \nn \\
\partial \tau_\lambda \bar\partial \tau_\lambda -
\tau_\lambda\partial\bar\partial\tau_\lambda =
2\lambda \tau_{\lambda - \2}^2; \nn \\
{\rm coefficient\ in\ front\ of\ }\bar\epsilon: \nn \\
2\lambda\tau_\lambda \bar\partial^2\tau_\lambda =
(2\lambda -1)(\bar\partial\tau_\lambda)^2; \nn \\
\ldots
\ee
If $\lambda = \frac{1}{2}$, the first one of these  is just
Liouville equation:
\be\label{Liouv}
\partial \tau_{\2} \bar\partial \tau_{\2} -
\tau_{\2}\partial\bar\partial\tau_{\2} =\tau_0^2 = 1,
\ee
or
\be
\partial\bar\partial \phi = 2e^{\phi}, \ \
\tau_{\2} = e^{-\phi/2},
\ee
while the second one,
\be\label{Virasoro}
\bar\partial^2\tau_{\2} = 0,
\ee
is very restrictive constraint.
Its role is to reduce the huge set of solutions to
Liouville equation,
\be\label{Louvsol}
\tau_{{\f 2}}(t,\bar t|g)=(1+A(t)B(\bar t))\left[{\partial
A\over\partial t}{\partial B\over\partial\bar t}\right]^{-\2},
\ee
parametrized by two {\it arbitrary} functions $A(t)$ and $B(\bar t)$,
to the 3-parametric family (\ref{3pfam}). In this sense it plays the same
role as the string equation (or Virasoro and $W$-constraints)
in matrix models. In the language of infinite-dimensional Grassmannian
there are infinitely many ways to embed $SL(2)$ group into
$GL(\infty)$ - and all the cases correspond to solutions (\ref{Louvsol})
(with some $A(t)$ and $B(\bar T)$) to the $SL(2)$ reduced
Toda-lattice hierarchy (i.e. Liouville equation), - but the constraint
(\ref{Virasoro})
specifies very concrete embedding: that in the left upper corner of
$GL(\infty)$ matrix (associated with {\it linear} functions
$A(t)$ and $B(\bar t)$).

\section{$\tau$-function for any $g \in SL_q(2)$}
\subsection{General solution of $SL_q(2)$ bilinear equations}

In the previous section we derived Hirota-like equations for the
$SL_q(2)$ $\tau$-functions. Formally these equations were derived
for any ``group element'' $g$, but so far we have explicitly
checked only that they are indeed satisfied by a $\tau$-fnction
at $g = I$, $\tau_\lambda(t,\bar t | g = I) = [ 1 + t\bar t]^{2\lambda}$.
It is of course easy to check this as well for any $g$ from
Cartan subgroup. However, restricting $g$ to Cartan subgroup,
we allow consideration of only (mutually) commuting matrix elements
$\langle {\bf k}| g |{\bf n}\rangle$ in (\ref{tau}).
The purpose of this section is to show explicitly that there is
no need in such restriction: the same equations are satisfied
for arbitrary choice of matrix elements from the non-commutative
coordinate ring $A(G)$.\footnote{To our knowledge,
the idea to study non-commutative
$\tau$-functions was first proposed in \cite{GS}.}
This result, when compared with the
derivation of Hirota equations, implies that the set of the
``group elements'', i.e. solutions to
$\Delta(g) = g\otimes g$, should be larger than Cartan
subgroup. This is indeed true, if we consider $g$ as belonging
to $U_q(G|A(G))$, which is defined as the universal enveloping
algebra (i.e. the set of formal series in generators $T_\pm$
and $q^{\pm T_0}$) with coefficients in $A(G)$.
Unfortunately we are not aware of any explicit construction of this
type and present just an illustrative
example in section 4.2.

Let us begin with the case of $\lambda = \2$. Then
\be
\tau_{\2}(t,\bar t|g) =
\langle + | g | + \rangle  + \bar t \langle + | g | - \rangle
+ t \langle - | g | + \rangle + t\bar t \langle - | g | - \rangle =\nn \\
= a + b\bar t + ct + dt\bar t,
\ee
where  $a,b,c,d$ are elements of the matrix
\beq
{\cal T} = \left(
\begin{array}{cc}
 a& b \\ c& d
\end{array}
\right)
\eeq
with the commutation relations dictated by ${\cal T}{\cal T}{\cal R}=
{\cal R}{\cal T}{\cal T}$ equation \cite{RTF}
\be
ab = qba, \nn \\
ac = qca, \nn \\
bd = qdb, \nn \\
cd = qdc, \nn \\
bc = cb, \nn \\
ad - da = (q - q^{-1})bc.
\label{core}
\ee
If $b$ or $c$ or both are non-vanishing, $\tau_{\2}(t,\bar t|g)$ with
different values of time-variables $t,\bar t$ do not commute.
Still such $\tau_{\2}(t,\bar t| g)$ does satisfy the same bilinear
identity (\ref{qeqA}), moreover, for this to be true it is essential that
commutation relations (\ref{core}) are exactly what they are.
Indeed, the l.h.s. of the equation (\ref{qeqA}) (using (\ref{1}) and
(\ref{core})) is equal to
\be\label{check}
-q^{\2}\sqrt{M^-_{\bar t}}(b+dt)\sqrt{M^+_{\bar t'}}(a+ct')+
q^{-\2}\sqrt{M^-_{\bar t}}(a+ct)\sqrt{M^+_{\bar t'}}(b+dt')=\\=
(q^{-\2}ab-q^{\2}ba)+(q^{-\2}cd-q^{\2}dc)tt'+(q^{-\2}cb-q^{\2}da)t+
(q^{-\2}ad-q^{-\2}bc)t'=\\=(q^{-\2}t'-q^{\2}t){\rm det}_qg,
\ee
which coincides with the r.h.s. of the equation (\ref{qeqA}).

To perform the similar check for any half-integer-spin representation,
let us note that the corresponding
$\tau$-function can be
easily written in terms of $\tau_{\2}$. Indeed,
\be
| n \rangle_{\lambda} = \left(q^{T_0}\otimes T_- +
T_- \otimes q^{-T_0}\right)^n| 0 \rangle_{\lambda -{\f 2}}\otimes
|0 \rangle_{\f 2} = \nn \\
= q^{-n/2}\left( |n \rangle_{\lambda-{\f 2}} \otimes |0\rangle_{\f 2} +
[n] q^{\lambda } |n-1\rangle_{\lambda -{\f 2}}\otimes | 1 \rangle_{\f 2}
\right); \\
\phantom._{\lambda}\langle n | =
\phantom._{\lambda-{\f 2}}\langle 0 | \otimes
\phantom._{\f 2} \langle 0 | =
\left(q^{T_0}\otimes T_+ + T_+\otimes q^{-T_0}\right)^n =
\nn \\ =
q^{-n/2} \left( \phantom._{\lambda-{\f 2}} \langle n | \otimes
\phantom._{\f 2}\langle 0 | + [n]q^{\lambda}
\phantom._{\lambda-{\f 2}} \langle n-1 | \otimes
\phantom._{\f 2}\langle 1 | \right).
\ee
Thus
\be
\phantom._{\lambda }\langle k | g | n \rangle_{\lambda} =
q^{-\frac{k+n}{2}}\left[
\phantom._{\lambda-{\f 2}} \langle k | g | n \rangle_{\lambda-{\f 2}}
\phantom. \langle + | g | + \rangle +
q^{\lambda }[n]\phantom._{\lambda-{\f 2}}\langle k | g | n-1
\rangle_{\lambda-{\f 2}} \langle + | g | - \rangle +\right.\\+ \left.
q^{\lambda }[k]\phantom._{\lambda-{\f 2}}\langle k-1 | g | n
\rangle_{\lambda-{\f 2}} \langle - | g | + \rangle +
q^{2\lambda}[k][n]\phantom._{\lambda-{\f 2}}\langle k-1 | g | n-1
\rangle_{\lambda-{\f 2}} \langle - | g | - \rangle \right]
\ee
or, in terms of generating ($\tau$-)functions:
\be
\tau_{\lambda }(t,\bar t|g) =
\sqrt{M^-_t M^-_{\bar t}}\left(\tau_{\lambda-\2}(t,\bar t|g)
\left(a+ q^{\lambda}\bar t b +
q^{\lambda}t c + q^{2\lambda } t\bar t d\right)\right).
\ee
Applying this procedure recursively we get:
\be
\tau_\lambda(t,\bar t|g) =
\tau_{\lambda - \2}(q^{-\2}t,q^{-\2}\bar t | g)
\tau_{\2}(q^{\lambda - \2}t,q^{\lambda - \2}\bar t | g) = \nn \\
\stackrel{{\rm if}\ \lambda \in \hbox{\bf Z}/2}{=}
\tau_{\2}(q^{\2 - \lambda}t, q^{\2 - \lambda}\bar t|g)
\tau_{\2}(q^{{3\over 2}-\lambda}t, q^{{3\over 2} - \lambda}\bar t | g) \ldots
\tau_{\2}(q^{\lambda - \2}t, q^{\lambda - \2}\bar t | g),
\ee
i.e. for half-integer $\lambda$ $\tau_\lambda$
is a polynomial of degree $2\lambda$ in $a,b,c,d$.

For example,
\be
\tau_1 (t,\bar t |g) = \tau_{\2}(q^{-\2}t,q^{-\2}\bar t | g)
\tau_{\2} (q^{\2}t, q^{\2}\bar t | g) = \nn \\
= (a + q^{-\2}\bar t b + q^{-\2} t c + q^{-1}t\bar t d)
(a + q^{\2}\bar t b + q^{\2}\bar t c + q t\bar t d) =
\nn \\ =
a^2 + (q^{\2}ab + q^{-\2}ba)\bar t +
(q^{\2}ac + q^{-\2}ca) t + b^2 \bar t^2 +\\+
(qad + bc + cb + q^{-1}da)t\bar t + c^2 t^2 +
(q^{\2}bd + q^{-\2}db)t\bar t^2 +\\+
 (q^{\2}cd + q^{-\2}dc)t^2\bar t+ d^2 t^2\bar t^2
\ee
Using the relations like
\be
q^{\2} ab + q^{-\2}ba = [2]q^{\2}ba = [2]q^{-\2}ab \ \ {\rm etc.}
\ee
one gets for this case
\be\label{tau=1}
\tau_1 (t,\bar t |g)=a^2+[2]q^{-\2}ab\bar t+[2]q^{-\2}act+b^2\bar
t^2+([2]qbc+[2]da)t\bar t+c^2t^2+ \nn \\ +
[2]q^{-\2}dbt\bar
t^2+[2]q^{-\2}dct^2\bar t+d^2t^2\bar t^2.
\ee
With this explicit expression, one can trivially make the calculations
similar to (ref{check})
in order to check manifestly equation (\ref{qeqA}) for $\lambda=1$,
$\lambda'={\2},\ 1$ and equation (\ref{qeqB}) for $\lambda= \lambda'={\2}$.

Thus, we showed explicitly (for the case of $SL_q(2)$) that the
quantum bilinear identities have as many solutions as the classical
ones, provided the $\tau$-function is allowed to take values in
non-commutative ring $A(G)$.

\subsection{Comment on the notion of ``group element''}

To avoid possible confusion we present here an illustrative example
of how the elements $g$ from the universal enveloping algebra
could look like, in order to guarantee that the ``group co-product''
$\Delta(g) = g\otimes g$ commutes with any intertwining operator.
In its turn the intertwining operators are defined as commuting
with the ``algebra coproduct'' $\Delta(T) \neq T\otimes I +
I\otimes T$.  Construction of such elements $g$ becomes possible
if $g$ is allowed to take values in $U_q(G|A(G))$: the
universal enveloping algebra with coefficients in
the non-commutative ``coordinate ring'' $A(G)$ (i.e. made out
of $a,b,c,d$-like objects). Unfortunately we know such $g$ only in a
form which strongly
depends on particular representation: more universal formulas
still remain to be found.

As everywhere in this paper, we restrict our example to the
simplest case: this time to a product of two spin-$\2$
representations of $U_q(SL(2))$, still this will be enough
to illustrate the most obscure points of the story.

So the claim is that the ``group element'' in a spin-$\2$
representation is equal to
\be
g = \alpha q^{T_0} + \beta q^{-T_0} + bT_+ + cT_-,
\ee
with
\be
\alpha = \frac{aq^{\2} - dq^{-\2}}{q-q^{-1}}, \ \ \ \
\beta = \frac{-aq^{-\2} + dq^{\2}}{q-q^{-1}}.
\ee
Then
\be
\Delta(g) = g\otimes g =
\alpha^2 q^{T_0}\otimes q^{T_0} + \beta^2 q^{-T_0}\otimes q^{-T_0}+\\
+ \alpha\beta q^{T_0}\otimes q^{-T_0} + \beta\alpha q^{-T_0}\otimes
q^{T_0} + bc (T_+\otimes T_- + T_-\otimes T_+) + \nn \\
+ (\alpha q^{T_0} + \beta q^{-T_0})\otimes bT_+ +
bT_+\otimes (\alpha q^{T_0} + \beta q^{-T_0}) + \nn \\
+ (\alpha q^{T_0} + \beta q^{-T_0})\otimes cT_- + cT_-\otimes
(\alpha q^{T_0} + \beta q^{-T_0}).
\ee
The r.h.s. does not look like any combination of
$\Delta(T_\pm) = q^{T_0}\otimes T_\pm + T_\pm\otimes q^{-T_0}$
and $\Delta(q^{nT_0}) = q^{nT_0}\otimes q^{nT_0}$ and it
can be a source of confusion. In fact, when acting on the
product of two spin-$\2$ representations, this expression
is just the same as
\be
\left((aq-d)\Delta(q^{T_0}) + (a-qd) \Delta(q^{-T_0}) -
(q+1)(a-d)\Delta(I)\right)\frac{(a-d)}{(q-1)(q-q^{-1})}
+\nn \\ +
(ad-qbc)\Delta(I) + \frac{bc}{[2]}
\left(q\Delta(T_+)\Delta(T_-) + q^{-1}\Delta(T_-)\Delta(T_+)\right)
+ \nn \\ +
\frac{q^{-\2}(a-d)b\Delta(q^{T_0}) - q^{\2}b(a-d)\Delta(q^{-T_0})}
{q-q^{-1}}\Delta(T_+) + \nn \\ +
\frac{q^{\2}(a-d)c\Delta(q^{T_0}) - q^{-\2}c(a-d)\Delta(q^{-T_0})}
{q-q^{-1}}\Delta(T_-)
\ee

   Once this identity is established, it is clear that $g$
satisfies all the necessary requirements. For it to be true, the actual
commutation relations between $a,b,c,d$ are of course important.

The simplest way to check the identity is just to see explicitly how
both expressions act on the four states $|+\rangle\otimes |+\rangle,\
|+\rangle\otimes|-\rangle, \ |-\rangle\otimes |+\rangle, \
|-\rangle\otimes |-\rangle$, and see that the action is the same.
A more instructive way is to observe that, in a given representation,
$q^{nT_0}$ with different $n$ are, in fact, {\it linearly}
dependent operators. In particular, in the spin-$\2$ representation
\be
q^{-T_0} = - q^{T_0} + s, \nn \\
q^{2T_0} = sq^{T_0} - 1, \nn \\
q^{-2T_0} = -sq^{T_0} + s^2 - 1
\ee
with $s = q^{\2} + q^{-\2}$ (for spin-$j$ representation there are
exactly $2j$ linearly independent operators $q^{nT_0}$ with integer $n$).
Using these relations together with
\be
q^{T_0}T_+ = q^{\2}T_+,
\ee
it is easy to prove the equivalence of two expressions  in a formal
way.

When higher-spin representations are considered, explicit
expressions for group elements are different (they are (homogeneous)
polynomials of degree $2j$ in $a,b,c,d$ and in $T_\pm$).
We are not yet aware of explicit universal formula (for all
representations at once, i.e. containing projectors, expressed
through $T$'s) even for the $U_q(SL(2))$ case, and it is an
interesting open problem to find it for all other groups.

\section{Fundamental representations}
\subsection{The set of fundamental representations of $SL(n)$}

This is another important example of our general construction,
which is the most close one to conventional case of integrable
hierarchies. The reason for this is that, in variance with
generic Verma modules for group $G \neq SL(2)$, the fundamental
representations are generated by subset of {\it mutually
commuting} operators, not by entire set of generators from maximal
nilpotent subalgebra. We describe the basic construction for $G = SL(n)$,
 since in this case
the (finite) Grassmannian construction is the most similar to the
conventional infinite-dimensional ($G = \hat U(1)$) situation.

There are as many as $r \equiv rank\ G = n-1$ fundamental
representations of $SL(n)$. They are the most conveniently described
in the language of Young tableaux, namely, in the following way.
Let us begin with the simplest fundamental representation $F$ -
the $n$-plet, consisting of the states
\be\label{simplfrep}
\psi_i = T_-^{i-1}| 0 \rangle_F, \ \ i = 1,\ldots,n.
\ee
Here the distinguished generator $T_-$ is essentially a sum of
those for all the $r$ {\it simple} roots of $G$:
$T_- = \sum_{i=1}^r T_{-{\bf\alpha}_i}$.
Then all the other fundamental representations $F^{(k)}$ are
defined as skew powers of $F = F^{(1)}$:
\be\label{frep}
F^{(k)} = \left\{ \Psi^{(k)}_{i_1\ldots i_k} \sim
 \psi_{[i_1}\ldots \psi_{i_k]} \right\}
\ee
{}From this description it is clear that $0 \leq k \leq n$,
moreover $F^{(0)}$ and $F^{(n)}$ are respectively the singlet and
dual singlet representations.
$F^{(k)}$ is essentially generated by the operators
\be\label{coprodFR}
R_k(T_-^i) \equiv T_-^i\otimes I \otimes \ldots \otimes I +
I\otimes T_-^i \otimes \ldots \otimes I +
I\otimes I \otimes \ldots \otimes T_-^i.
\ee
These operators commute with each other. It is clear that for
given $k$ exactly $k$ of them (with $i = 1,\ldots,k$) are
independent
(note that $R_k (T_-^i) \neq (\left(R_k(T_-)\right)^i$).

The intertwining operators which are of interest for us are
\be\label{inttwfrep}
I_{(k)}:\ \ F^{(k+1)} \longrightarrow F^{(k)} \otimes F, \nn \\
I^*_{(k)}:\ \ F^{(k-1)} \longrightarrow F^*\otimes F^{(k)},
\ \ {\rm and} \nn \\
\Gamma_{k|k'}:\ \ F^{(k+1)}\otimes F^{(k'-1)} \longrightarrow
F^{(k)} \otimes F^{(k')}.
\ee
Here
\be\label{frep2}
F^* = F^{(r)} = \left\{\psi^i \sim \epsilon^{ii_1\ldots i_r}
  \psi_{[i_1}\ldots \psi_{i_r]} \right\}, \nn \\
I_{(k)}:\ \ \Psi^{(k+1)}_{i_1\ldots i_{k+1}} =
            \Psi^{(k)}_{[i_1\ldots i_k}\psi^{\phantom{fgh}}_{i_{k+1}]}, \nn \\
I^*_{(k)}:\ \ \Psi^{(k-1)}_{i_1\ldots i_{k-1}} =
            \Psi^{(k)}_{i_1\ldots i_{k-1}i}\psi^i,
\ee
and $\Gamma_{k|k'}$ is constructed with the help of embedding
$I \longrightarrow F \otimes F^*$, induced by the pairing
$\psi_i \psi^i$: the basis in linear space $F^{(k+1)}\otimes
F^{(k'-1)}$, induced by $\Gamma_{k|k'}$ from that in
$F^{(k)}\otimes F^{(k')}$ is:
\be\label{prodfrep}
\Psi^{(k)}_{[i_1\ldots i_k}\Psi^{(k')}_{i_{k+1}]i'_1\ldots i'_{k'-1}}.
\ee
Operation $\Gamma$ can be now rewritten in terms of matrix elements
\be\label{gkdet}
g^{(k)}\left({{i_1\ldots i_k}\atop{j_1\ldots j_k}}\right) \equiv
\langle \Psi_{i_1\ldots i_k} | g | \Psi_{j_1\ldots j_k} \rangle=
\det_{1\leq a,b\leq k} g^{i_a}_{j_b}
\ee
as follows:
\be\label{gkgk}
g^{(k)}\left({{i_1\ldots i_k}\atop{[j_1\ldots j_k}}\right)
g^{(k')}\left({{i'_1\ldots i'_k}\atop
{j_{k+1}]j'_1\ldots j'_{k'-1}}}\right) = \nn \\ =
g^{(k+1)}\left({{i_1\ldots i_k[i'_{k'}}
\atop{j_1\ldots j_{k+1}}}\right)
g^{(k'-1)}\left({{i'_1\ldots i'_{k'-1}]}\atop
{j'_1\ldots j'_{k'-1}}}\right)
\ee
This is the explicit expression for eq.(\ref{CRGamma}) in the case of
fundamental representations, and it is certainly identically true for any
$g^{(k)}$ of the form (\ref{gkdet}).\footnote{
To see this directly it is enough to rewrite the l.h.s. of
(\ref{gkgk}) as
$$
g^{(k)}\left({{i_1\ldots i_k}\atop{[j_1\ldots j_k}}\right)
g^{[i_k'}_{j_{k+1}]}
g^{(k'-1)}\left({{i'_1\ldots i'_{k-1}]}\atop{j'_1\ldots j'_{k-1}}}\right)
$$
(expansion of the determinant $g^{(k')}$ in the first column) and now
 the first two factors can be composed
into $g^{(k+1)}$ (expansion of the determinant $g^{(k+1)}$ in the first
row), thus giving the r.h.s. of (\ref{gkgk}).}

Let us note that one can easily construct from the minors (\ref{gkdet})
local coordinates in the Grassmannian \cite{Plucker}. Then, these
coordinates satisfy a set of (bilinear) Plucker
relations \cite{Plucker}, which are nothing but defining equations of
the Grassmannian consisting of all $k$-dimensional vector subspaces of
an $n$-dimensional vector space. Parametrizing determinants (\ref{gkdet})
by time variables (see (\ref{H})), one get a set of
bilinear differential equations on the generating function of these
Plucker coordinates, which is just $\tau$-function \cite{Ohta}.

Now let us introduce time-variables and rewrite (\ref{gkgk}) in terms
of $\tau$-functions. We shall denote time variables through
$s_i, \bar s_i$, $i = 1,\ldots,r$ in order to emphasize their
difference from generic $t_{\bf\alpha}, \bar t_{\bf\alpha}$
labeled by all the positive roots ${\bf\alpha}$ of $G$. Note that in
order to have a closed system of equations we need to introduce all the
$r$ times $s_i$ for all $F^{(k)}$ (though $\tau^{(k)}$ actually depends
only on $k$ independent combinations of these).

Since the highest weight of representation $F^{(k)}$ is
identified as
\be
| 0 \rangle_{F^{(k)}} = |\Psi^{(k)}_{1\ldots k} \rangle,
\ee
we have:
\be\label{tau-k}
\tau^{(k)}(s,\bar s\ |\ g) =
\langle \Psi^{(k)}_{1\ldots k} |
\exp \left(\sum_i s_i R_k(T_+^i)\right)\ g \
\exp\left(\sum_i \bar s_iR_k(T_-^i)\right) |
\Psi^{(k)}_{1\ldots k} \rangle .
\ee

Now,
\be\label{70}
\exp\left(\sum_i s_i R_k(T^i)\right) =
\exp\left(R_k\left(\sum_i s_iT^i\right)\right) = \nn \\
= \left( \exp\left(\sum_i s_i T^i\right)\right)^{\otimes k} =
\left( \sum_j P_j(s)T^j\right)^{\otimes k},
\ee
where we used the definition of the Schur polynomials
\be\label{Schur}
\exp\left(\sum_i s_iz^i\right) = \sum_j P_j(s)z^j,
\ee
their essential property being:
\be\label{SS}
\partial_{s_i} P_j(s) = (\partial_{s_1})^i P_j(s) = P_{j-i}(s).
\ee
Because of (\ref{70}), we can rewrite the r.h.s. of (\ref{tau-k}) as
\be\label{detrep}
\tau^{(k)}(s,\bar s\ |\ g) =
\nn \\
= \sum_{{i_1,\ldots,i_k}\atop{j_1,\ldots,j_k}}
P_{i_1}(s)\ldots P_{i_k}(s) \langle \Psi^{(k)}_{1+i_1,2+i_2,\ldots,k+i_k}
|\ g\ | \Psi^{(k)}_{1+j_1,2+j_2,\ldots,k+j_k} \rangle
P_{j_1}(\bar s)\ldots P_{j_k}(\bar s) = \nn \\ =
\det_{1\leq \alpha,\beta \leq k} H^\alpha_\beta(s,\bar s),
\ee
where
\be\label{H}
H^\alpha_\beta(s,\bar s) = \sum_{i,j} P_{i-\alpha}(s)g^i_j
P_{j-\beta}(\bar s).
\ee
This formula can be considered as including infinitely many
times $s_i$ and $\bar s_i$, and it is only due to the finiteness
of matrix $g^i_j \in SL(n)$ that $H$-matrix is additionally constrained
\be\label{excon}
\left(\frac{\partial}{\partial s_1}\right)^n H^\alpha_\beta = 0, \nn \\
\ldots \nn \\
\frac{\partial}{\partial s_i}H^\alpha_\beta = 0, \ \ {\rm for}\ \
i \geq n.
\ee
The characteristic property of $H^\alpha_\beta$ is that it satisfies
the following ``shift'' relations (see (\ref{SS})):
\be\label{der}
\frac{\partial}{\partial s_i}H^\alpha_\beta  = H^{\alpha +i}_\beta, \ \ \
\frac{\partial}{\partial \bar s_i} H^\alpha_\beta =
H^\alpha_{\beta + i}.
\ee
Expressions (\ref{detrep}), (\ref{H}) and (\ref{der})
are, of course, familiar from the theory of KP
and Toda hierarchies (see \cite{Mamo2,AM}
and references therein).

Coming back to bilinear relation (\ref{gkgk}), it can be easily
rewritten in terms of $H$-matrix: it is enough to convolute them
with Schur polynomials. For the sake of convenience let us denote
$H\left({\alpha_1\ldots\alpha_k}\atop{\beta_1\ldots \beta_k}\right)
= \det_{1\leq a,b \leq k} H^{\alpha_a}_{\beta_b}$. In accordance with
this notation
$\tau^{(k)} = H\left({1\ldots k}\atop{1\ldots k}\right)$, while bilinear
equation turns into:
\be
H\left({\alpha_1\ldots \alpha_k}\atop{[\beta_1\ldots \beta_k}
\right)
H\left({\alpha'_k\alpha'_1\ldots \alpha'_{k-1}}\atop
{\beta_{k+1}]\beta'_1\ldots\beta_{k-1}'}\right) =
H\left({\alpha_1\ldots \alpha_k[\alpha'_k}\atop
{\beta_1\ldots\beta_k\beta_{k+1}}
\right)
H\left({\alpha'_1\ldots\alpha_{k-1}]'}\atop{\beta'_1\ldots\beta'_{k-1}}\right).
\ee
Just like original (\ref{gkgk}) these are just matrix identities, valid for
any $H^\alpha_\beta$. However, after the switch from $g$ to $H$
we, first, essentially represented the equations in $n$-independent
form and, second, opened the possibility to rewrite them in terms
of time-derivatives.

For example, in the simplest case of
\be
\alpha_i = i, \ \ i = 1,\ldots, k'; \nn \\
\beta_i = i, \ \ i = 1,\ldots,k+1; \nn \\
\alpha'_i = i, \ \ i = 1,\ldots,k-1,\ \ \alpha'_k = k+1; \nn \\
\beta'_i = i, \ \ i = 1,\ldots,k-1
\ee
we
get:
\be
H\left({1\ldots k}\atop{1\ldots k}\right)
H\left({k+1,1\ldots k-1}\atop{k+1,1\ldots k-1}\right) -\\-
H\left({1\ldots k-1, k}\atop{1\ldots k-1, k+1}\right)
H\left({k+1,1\ldots k-1}\atop{k,1\ldots,k-1}\right) =\\=
H\left({1\ldots k+1}\atop{1\ldots k+1}\right)
H\left({1\ldots k-1}\atop{1\ldots k-1}\right)
\ee
(all other terms arising in the process of symmetrization vanish).
This in turn can be represented through $\tau$-functions:
\be\label{hirotafr}
\partial_1\bar\partial_1\tau^{(k)} \cdot \tau^{(k)} -
\bar\partial_1\tau^{(k)} \partial \tau^{(k)} =
\tau^{(k+1)}\tau^{(k-1)}.
\ee
This is the usual lowest Toda-lattice equation. For finite $n$
the set of solutions is labeled by $g \in SL(n)$ as a result of
the additional constraints (\ref{excon}).

   We can now use the chance to illustrate the ambiguity of definition
of $\tau$-function, or, to put it differently,
that in the choice of time-variables.
Eq.(\ref{hirotafr})
is actually a corollary of {\it two} statements: the basic identity
(\ref{gkgk}) and the particular definition (\ref{tau}), which in
this case implies (\ref{H}) with $P$'s being ordinary
Schur polynomials (\ref{Schur}). At least, in this simple situation
(of fundamental representations of $SL(n)$) one could define
$\tau$-function not by eq.(\ref{tau}), but just by eq.(\ref{detrep}), with
\be
H^\alpha_\beta(s,\bar s) \longrightarrow
{\cal H}^\alpha_\beta(s,\bar s) = \sum_{i,j} {\cal P}_{i-\alpha}(s)\
g^i_j\ {\cal P}_{j-\beta}(\bar s)
\ee
with {\it any} set of independent functions (not even polynomials)
${\cal P}_\alpha$. Such
\be
\tau^{(k)}_{{\cal P}} = \det_{1\leq \alpha,\beta \leq k}
{\cal H}^\alpha_\beta
\ee
still remains a generating function for all matrix elements of $G=SL(n)$
in representation $F^{(k)}$. This freedom should be kept in
mind when dealing with ``generalized $\tau$-functions''. As a simple
example, one can take ${\cal P}_\alpha(s)$ to be $q$-Schur polynomials,
\be
\prod_i e_q(s_iz^i) = \sum_j P^{(q)}_j(s)z^j,  \ \ {\rm or} \nn \\
\prod_i e_{q^i}(s_iz^i) = \sum_j \hat P^{(q)}_j(s)z^j,
\ee
which satisfy (hereafter we denote $D\equiv D^{(0)}$)
\be
D_{s_i} P^{(q)}_j(s) = (D_{s_1})^i P^{(q)}_j(s) = P^{(q)}_{j-i}(s).
\ee
Then instead of (\ref{der})
we would have:
\be
D_{s_i}{\cal H}^\alpha_\beta = {\cal H}^{\alpha +i}_\beta, \ \
D_{\bar s_i}{\cal H}^\alpha_\beta = {\cal H}^\alpha_{\beta+i}
\ee
and
\be
\tau_{P^{(q)}}^{(k)} (s,\bar s | g) =
\det_{1\leq \alpha,\beta \leq k} D_{s_1}^{\alpha -1}
D_{\bar s_1}^{\beta -1} {\cal H}^1_1(s,\bar s).
\ee
So defined $\tau$-function satisfies {\it difference} rather than
differential equations \cite{Sat,MMV}:
\be
\tau^{(k)}\cdot D_{s_1}D_{\bar s_1}\tau^{(k)}-
D_{s_1}\tau^{(k)}\cdot D_{\bar
s_1}\tau^{(k)}=\tau^{(k-1)}\cdot M^+_{s_1}M^+_{\bar s_1}\tau^{(k+1)},\\
\ldots\ \ \ .
\ee
We emphasize, however, that, in a sense,
this is just a redefinition of the
$SL(n)$, not $SL_q(n)$ $\tau$-function, as generating function of matrix
elements. In particular, this $\tau$-function
is a $c$- rather than $q$-number function. Still it would have something
to do with $SL_q(n)$ group, but as a function of times, i.e. rather in
spirit of the connection of
$q$-hypergeometric functions to quantum groups (see, for example,
\cite{Vinet}).

\subsection{Approach to $SL_q(n)$}

In order to extend this reasoning to the case of $SL_q(n)$ with
$q\neq 1$ we need to go into some more details about the group
structure.

The main thing we shall need is the notion of $q$-antisymmetrization,
to be defined as a sum over all perturbations,
\be
\left( [1,\ldots,k]_q\right) = \sum_P (-q)^{{\rm deg}\ P}
\left(P(1),\ldots,P(k)\right),
\ee
where
\be
{\rm deg}\ P = \#\ {\rm of\ inversions\ in}\ P.
\nn
\ee
The first place to use this notion is the definition of $q$-determinant:
\be
{\rm det}_q A \sim  A^{[1}_{[1}\ldots A^{n]_q}_{n]_q}
= \sum_{P,P'} (-q)^{{\rm deg}\ P + {\rm deg}\ P'}
\prod_{a} A^{P(a)}_{P'(a)}.
\label{fqdet}
\ee
Note that this is not necessarily the same as $A^1_{[1}\ldots A^n_{n]_q}$,
for example for $n=2$ (\ref{fqdet}) gives
$\displaystyle{\frac{1}{[2]}(A^1_1A^2_2 - qA^1_2A^2_1 -
qA^2_1A^1_2 + q^2A^2_2A^1_1)}$, while
single $q$-antisymmetrization would give just
$A^1_1A^2_2 - qA^1_2A^2_1$. Moreover,
$A^1_{[1}A^1_{2]} = A^1_1 A^2_2 - q A^1_2 A^1_1$ does not need to vanish,
and even $A^1_{[1}A^1_{1]_q} = (1-q)(A^1_1)^2 \neq 0$.

The ``normal'' properties of $q$-antisymmetrization are restored only when
$A$ is considered as being an element of $GL_q(n)$.
This means
that its elements take values in the non-commutative ring
$A(GL(n))$ and the following commutation relations - essentially
the same as (\ref{core}) - are imposed \cite{RTF}:
\be
\forall i, \forall j_1<j_2 \ \ \ \
A^i_{[j_1}A^i_{j_2]_q} = 0, \ \ \ \ (ab=qba,\ cd=qdc) \nn \\
\forall i_1<i_2, \forall j \ \ \ \
A^{[i_1}_jA^{i_2]_q}_j = 0, \ \ \ \ (ac = qca,\ bd = qdb) \nn \\
\forall i_1\neq i_2, j_1\neq j_2 \ \ \ \
A^{i_1}_{j_2}A^{i_2}_{j_1} = A^{i_2}_{j_1}A^{i_1}_{j_2},
 \ \ \ \ (bc = cb) \nn \\
\forall i_1<i_2,\ j_1<j_2 \ \ \ \\
A^{i_1}_{j_1} A^{i_2}_{j_2} - A^{i_2}_{j_2}A^{i_1}_{j_1} =
(q - q^{-1}) A^{i_1}_{j_2}A^{i_2}_{j_1} \ \ \ \
(ad-da = (q-q^{-1})bc).
\ee
For $A \in GL_q(n)$
\be
{\rm det}_q A = A^1_{[1}\ldots A^n_{n]_q} = A^{[1}_1\ldots A^{n]_q}_n.
\ee
and $A^{i_1}_{[1}\ldots A^{i_k}_{k]_q} = 0$, if any two of the
upper indices coincide (but only provided the lower ones are all
[6~different: it is still true that $A^{1}_{[1}A^{1}_{1]_q} =
(1-q)(A^1_1)^2 \neq 0$ even for $A \in GL_q(n)$).

The notion of $q$-antisymmetrization is important for us
because the $k$-th fundamental representation $F^{(k)}$ of
$SL_q(n)$ is the $q$-skew power of $F = F^{(1)}$: for $q\neq 1$
we have instead of (\ref{frep}):
\be
F^{(k)} = \left\{\Psi^{(k)}_{i_1\ldots i_k} \sim
\psi_{[i_1}\ldots \psi_{i_k]_q}, \ \ \ i_1<\ldots <i_k\right\}
\label{fqrep}
\ee
Now it is necessary to request {\it explicitly} that all $i_a$
are different.

All the formulas (\ref{inttwfrep})--(\ref{prodfrep}) for intertwining
operators remain exactly the same with antisymmetrization replaced
by $q$-antisymmetrization (and obvious definition of $q$-$\epsilon$-symbol).
Instead of (\ref{gkdet}) we have:
\be\label{gkdetq}
g^{(k)}\left({i_1\ldots i_k}\atop{j_1\ldots j_k}\right) \sim
{\rm det}_q g^{i_a}_{j_b}, \ \ \ \ i_1 < \ldots i_k,\ \hbox{or}\
j_1<\ldots j_k,
\ee
and (\ref{gkgk}) turns into:
\be
g^{(k)}\left({i_1\ldots i_k}\atop{[j_1\ldots j_k}\right)
g^{(k')}\left({i'_1\ldots i'_{k'}}\atop{j_{k+1}]_q j'_1\ldots j'_{k'-1}
}\right) = \\=
g^{(k+1)}\left({i_1\ldots i_k[i'_k}\atop{j_1\ldots j_{k+1}}\right)
g^{(k'-1)}\left({i'_1\ldots i'_{k'-1}]_q}\atop{j'_1\ldots j'_{k'-1}}
\right),
\label{gqgq}
\ee
\be
i_1 <\dots< i_k, \ \ \hbox{or}\ \ j_1<\ldots j_k,\ \ \ \ \
i_1'<\ldots<i_{k'}'\ \ \hbox{or}\ \ j_{k+1}<j_1'<\ldots<j_{k'-1}',\\
i_1<\ldots<i_k<i'_{k'},\ \hbox{or}\ j_1<\ldots<j_{k+1},\ \ \
i_1'<\ldots<i'_{k'-1},\ \hbox{or}\ j'_1<\ldots<j'_{k'-1}.
\ee
Just as (\ref{gkgk}) this is nothing but an identity for the
matrices from $GL_q(n)$.\footnote{It is crucially
important here that $g$ indeed belongs to some $GL_q(n)$, i.e. its
elements have the proper commutation relations. This is of course
{\it implied} by the derivation of (\ref{gqgq}), where $g$ is
supposed to be a ``group element''. Be it not the case, we would
need to understand (\ref{gkdetq}) in the sense of (\ref{fqdet})
(in particular, to write {\it and} instead of {\it or} in (\ref{gkdetq}))
and then we would run in a contradiction with (\ref{gqgq}).
To make it more transparent, let us take $k = k' = 1$,
then (\ref{gqgq}) becomes:
\be
g^{i}_{[j_1}g^{i'}_{j_2]_q} = g^{(2)}\left({i\ i'}\atop{j_1j_2}\right)
\nn
\ee
and the l.h.s. is equal to $g^i_{j_1}g^{i'}_{j_2} - q
g^{i}_{j_2}g^{i'}_{j_1}$, while the r.h.s. would be interpreted as
$g^{i}_{j_1}g^{i'}_{j_2} - q g^{i}_{j_2}g^{i'}_{j_1} -
q g^{i'}_{j_1}g^i_{j_2} + q^2g^{i'}_{j_2}g^i_{j_1}$.}
A feature which is essentially new as compared to (\ref{gkgk})
is explicit appearance of restrictions on indices $i,j,i',j'$,
which makes the translation to the language of generating
functions a little more sophisticated.

Let us note that, similar to the classical case, one can construct from the
quantum minors (\ref{gkdetq}) local coordinates on the quantum flag
space \cite{Noumi}. As before, in the quantum case there is a set of
(quantum) bilinear Plucker relations. Unfortunately, the problems arise
when parametrizing Plucker coordinates by time variables.

Just to give an impression of what the result can be, when
somehow expressed in terms of $H$, let us restrict ourselves
to the case of $G = SL_q(2)$ and introduce:
\be
H^1_1 = \tau_F = a + b\bar t + ct + dt\bar t.
\ee
If
\be
H^1_2 = D_{\bar t}H^1_1 = b+dt, \nn \\
H^2_1 = D_t H^1_1 = c+d\bar t, \nn \\
H^2_2 = D_{\bar t}D_t H^1_1 = d,
\ee
we see that $H^a_b$ is actually not lying in $GL_q(2)$
(for example,
$H^1_2 H^2_1 \neq H^2_1 H^1_2$), i.e. a matrix consisting of
the $\tau_F$ and its derivatives, despite these are all
elements of $A(G)$, does not longer belong to $G_q$. Thus, it is not
reasonable to consider ${\rm det_q}H$ (or the definition of $H$ should
be somehow modified). Instead the appropriate
formula for the case of $SL_q(2)$ looks like
\be
\tau_{F^{(2)}} = {\rm det}_q g =
H^1_1 H^2_2 - q H^1_2 M^-_{\bar t} H^2_1 =
\tau_F D_tD_{\bar t}\tau_F - q D_{\bar t}\tau_F M^-_tD_t\tau_F.
\ee

We shall not go into more discussion of transition from $g$-identity
to $H$-identities, because it involves some art in the work with
appropriate time-variables, and is not yet brought to a reasonably
simple form. Instead we present a few more formulas, which can be
illuminating for some readers.

\subsection{Comments on the quantum case}

The first thing we wish to give some more details on is the statement
(\ref{fqrep}).

The Lie algebra $SL(n)$ is generated by operators $T_{\pm{\bf\alpha}}$
and Cartan operators $H_{\bf\beta}$, such that
$[H_\beta, T_{\pm{\bf\alpha}}] = \pm\2({\balpha\bbeta})
T_{\pm{\bf\alpha}}$. All elements of all representations are
eigenfunctions of $H_{\bf\beta}$, $H_{\bf\beta}|\blambda\rangle
= \2({\bbeta\blambda})|\blambda\rangle$. The highest weight
of representation $F^{(k)}$ is ${\bmu}_k$. Vectors
${\bmu}_k$'s
are ``dual'' to the {\it simple} roots ${\balpha}_i$, $i = 1,\ldots,r$:
$({\bmu}_i{\balpha}_j) = \delta_{ij}$, and
${\brho} = \frac{1}{2}\sum_{{\bf\alpha}>0}{\balpha} =
\sum_i {\bmu}_i$.

Representation $F^{(1)}$ consists of the states
\be
\psi_i = T_{-(i-1)}\ldots T_{-2}T_{-1}\psi_1, \ \ \ i = 1,\ldots,n.
\label{psii}
\ee
Moreover
\be
T_{-i}\psi_j = \delta_{ij}\psi_{i+1}
\ee
(thus, for $T_- = \sum_{i=1}^r T_-i\ $  $T_-^i\psi_j = \psi_{j+i}$ and
(\ref{simplfrep}) follows),
and
\be
\blambda(\psi_i) = {\bmu}_1 - {\balpha_1} - \ldots -
{\balpha}_{i_1}.
\ee
Here $T_{\pm i} \equiv T_{\pm{\bf\alpha}_i}$ are generators, associated
with the simple roots. Let us denote the corresponding basis in
Cartan algebra $H_i = H_{{\bf\alpha}_i}$, and
${H_i}|{\blambda}\rangle = \2({{\balpha}_i{\blambda}})
|{\blambda}\rangle = \lambda_i|{\blambda}\rangle$.
Then
\be\label{lambda}
\lambda_i^{(j)} \equiv \lambda_i(\psi_j) = {\f 2}(\delta_{ij} -
\delta_{i,j-1}).
\ee
This formula, together with (\ref{psii}) and (\ref{notations}) implies that
$||\psi_i||^2 = 1$, and, since comultiplication formula in the
classical case is just $\Delta(T) = T\otimes I + I\otimes T$,
it is obvious that $\psi_{[1}\ldots \psi_{k]}$ are all highest
weight vectors (i.e are annihilated by all $\Delta_k(T_{+i})$ and,
thus by all the $\Delta_k(T_{+{\bf\alpha}})$).

   Quantum universal enveloping algebra $U_q(SL(n))$ is generated by
$T_{\pm i}$ and $q^{\pm H_i}$ with basic commutation relations ($a_{ij}$
is the Cartan matrix of $SL(n)$)
\be\label{notations}
q^{H_i} T_{\pm j} q^{-H_i} = q^{\pm a_{ij}}T_{\pm j}, \nn \\
\phantom. [T_{+i},T_{-j}] = \delta_{ij}\frac{q^{2H_i}-q^{-2H_i}}{q - q^{-1}}
\ee
and comultiplication law
\be
\Delta(T_{\pm i}) = q^{H_i}\otimes T_{\pm i} + T_{\pm i}\otimes
q^{-H_i}, \nn \\
\Delta(q^{\pm H_i}) = q^{\pm H_i}\otimes q^{\pm H_i}.
\ee
Comultiplication formulas for $T_{\pm\alpha}$ in the case of non-simple
roots are corollaries of these and look more sophisticated. For
example, for the ``height 2'' ${\balpha}$, such that
$T_{\pm{\bf\alpha}} = \pm [ T_{\pm {\bf\alpha}_i}, T_{\pm{\bf\alpha}_{i+1}}]$,
we have
\be
\Delta(T_{-{\bf\alpha}}) = - [\Delta(T_{\alpha_i}),\Delta(T_{\alpha_{i+1}})]
= q^{H_\alpha}\otimes T_\alpha + T_\alpha \otimes q^{-H_\alpha} +\\+
(q^{1/2}-q^{-1/2})\left[(T_{-\alpha_i}\otimes T_{-\alpha_{i+1}})
(q^{H_{i+1}}\otimes q^{-H_i}) -
(T_{-\alpha_{i+1}}\otimes T_{-\alpha_i})(q^{H_i}\otimes q^{-H_{i+1}})\right].
\ee

Given multiplication formulas, one can easily check that indeed (\ref{fqrep})
is true. For example, for $F^{(2)}$:
\be
\Delta(T_{+i}) (\psi_1\psi_2 - q\psi_2\psi_1) =
\delta_{i,1}(q^{\lambda_1^{(1)}}\psi_1\psi_1 - q^{1-\lambda_1^{(1)}}
\psi_1\psi_1) = 0,
\ee
because $\lambda_1^{(1)} = \frac{1}{2}$. Thus $\Psi^{(2)}_{12} \equiv
\psi_{[1}\psi_{2]_q}$ is indeed the highest weight vector.
Similarly $(i<j)$:
\be
\Delta(T_{-l})\Psi^{(2)}_{ij} = \Delta(T_{-l})(\psi_i\psi_j - q\psi_j
\psi_i) = \nn \\
= \delta_{li} q^{-\lambda_i^{(j)}}(\psi_{i+1}\psi_j -
q^{1+2\lambda_i^{(j)}}\psi_j\psi_{i+1}) +
  \delta_{lj} q^{\lambda_j^{(i)}}(\psi_i\psi_{j+1} -
q^{1-2\lambda_j^{(i)}}\psi_{j+1}\psi_i).
\ee
According to (\ref{lambda}) for $i<j$ $\lambda_j^{(i)} = 0$ in all cases,
while $\lambda_i^{(j)}\neq 0$ only if $j = i+1$, and
$\lambda_i^{(i+1)} = -{\f 2}$. Thus,
\be
\Delta(T_{-l})\Psi_{ij}^{(2)} = \delta_{lj}
\Psi^{(2)}_{i,j+1} + \delta_{li}\Psi^{(2)}_{i+1,j}(1 - \delta_{i+1,j}).
\ee
The rules for the action of all $\Delta(T_{-\alpha})$ follow from this.
It is easy to describe explicitly the action of all
$\Delta(T_{+i})$ and also to do the same for all other representations
$F^{(k)}$.

Our second comment concerns the relevance of $q$-exponents in eq.(\ref{tau})
in the quantum group case. Of course, they are primarily needed in
order to obtain the Hirota equations in the nice form of difference
equations. This, however, does not fix the choice completely.
Indeed, there are various ways to define $q$-numbers and thus $q$-derivatives
and $q$-special functions. These ways are in correspondence with the
choices of comultiplication in quantum algebra. Through this text
we are using ``symmetric'' coproduct (\ref{coprod}), thus $q$-numbers
need to be defined in symmetric way as well. For given definition of
$q$-numbers there is still an ambiguity in the choice of $q$-exponent.
In (\ref{tau}) and in section 3 we used the simplest
definition\footnote{One can also keep in mind the possibility to
consider ${\balpha}$-dependent $q$ in (\ref{tau}) (see for example
\cite{MMV}). The reasonable choices of $q_\alpha$ could be
$q_\alpha = q^{{\bf\alpha}^2/2}$ or $q_\alpha = q^{||{\balpha}||}$,
where $||{\balpha}||$ is the ``height'' of the root ${\balpha}$,
$||{\balpha}|| = {\balpha\brho}$.}
\be
e_q(x) = \sum \frac{x^n}{[n]!}, \ \ {\rm satisfying}\ \
D_xe_q(x) = e_q(x).
\ee
However, it can appear more reasonable in some situations to use
more sophisticated choices, for example,
\be
E_q(x) = \sum_n q^{n(n-1)/2}\frac{x^n}{[n]!}, \ \
{\rm satisfying}\ \ D_xE_q(x) = M^+_xE_q(x) = E_q(qx).
\ee
The advantage of such $q$-exponent is that it satisfies the
``summation rule''
\be
E_q(x) E_q(y) = E_q(x+y),\ \ {\rm for}\ \ xy = q^2yx.
\ee
With this definition we have, for example, the following ``decoupling''
property:
\be
E_q(\bar t_i\Delta_k(T_{-i})) =
E_q(\bar t_i T_{-i}\otimes q^{-H_i}\otimes \ldots \otimes
q^{-H_i})\times
E_q(\bar t_i q^{H_i}\otimes T_{-i}\otimes \ldots \otimes
q^{-H_i})\times\\\times
\ldots\times
E_q(\bar t_i q^{H_i}\otimes q^{H_i}\otimes \ldots T_{-i})
\ee
for each $i = 1,\ldots r$ (but not for the sum over $i$).

\section{Conclusion}

To summarize, we suggest to begin investigation of the concept
of $\tau$-function in the very general framework, defining
it as a generating function for all the matrix elements of
a group in a given representation. In physical language this
notion is of course very close to that of the ``non-perturbative
partition function'', which contains information about all the amplitudes
in the theory. We argued that the bilinear Hirota-like identities,
mixing various representations,
are a very general feature of such generating functions, as
are the analogues of Virasoro constraints and ``string
equation''. This opens the road to a very general group-theoretical
interpretation of non-perturbative partition functions in
quantum mechanics, field and string theories.

As a byproduct of this general formalism we derived an example
of Hirota equations for quantum groups, which incorporate
properly the non-commutative ($q$-number) nature of the
corresponding $\tau$-functions.
   This $q$-number $\tau$-function can be further transformed into
a $c$-number one in a particular representation of coordinate ring
$A(G)$. Such procedure is of course reminiscent of the general
ideology of the second quantized string theory (when the partition
function of the first-quantized model is further substituted into
the functional integral over the space of theories).

The most important example which remains to be discussed in order to
demonstrate all the new features of ``generalized $\tau$-functions''
is the case of non-fundamental representations of $G=SL(3)$,
where for the first time the existence of ``non-Cartanian''
time-variables will be essential. As to the fundamental
representations of $G=SL(n)$, there exists a closed
subsystem of Hirota equations for them, which does not require
more that $n-1$ time-variables, and all of these
can be associated with commuting generators. This example
is described in some detail in section 5, though its quantum
analogue is to be worked out in more details. In particular, the
interpretation within the terms of quantum flag space (in spirit of
\cite{Noumi}) is to be obtained in the proposed framework.

\section*{Acknowledgements}

We are indebted to S. Kharchev and A. Zabrodin for
numerous discussions. A.Mironov is grateful to the Niels Bohr Institute,
and especially Jan Ambjorn for the kind hospitality and support.
A.Morozov acknowledges
hospitality and support of the Volterra Center at Brandeis University.
The work of A.Gerasimov and A. Mironov is
partially supported by grant 93-02-14365 of the Russian
Foundation of Fundamental Research.

\end{document}